\documentstyle[11pt,newpasp,twoside,epsf]{article}
\def\edcomment#1{\iffalse\marginpar{\raggedright\sl#1\/}\else\relax\fi}
\reversemarginpar

\begin{document}
\title{The Chemical Composition of Carbon C(N) stars}
 \author{C. Abia, I. Dom\'\i nguez}
\affil{Dpt. F\'\i sica Te\'orica y del Cosmos. Universidad de Granada. 18071
Granada. Spain}
\author{R. Gallino, S. Masera}
\affil{Dipartimento di Fisica Generale. Universit\'a di Torino. 10125 Torino. Italy}
\author{M. Busso}
\affil{Dipartimento di Fisica. Universit\'a di Perugia. 06123 Perugia. Italy}
\author{O. Straniero}
\affil{Osservatorio di Collurania. 64100 Teramo. Italy}
\author{P. de Laverny}
\affil{Observatoire de la Cote d'Azur. 06528 Nice. France}
\author{B. Plez}
\affil{GRAAL. Universit\'e de Montpellier II. Montpellier. France}

\begin{abstract}
A chemical study of normal Galactic C(N) carbon
stars is presented. Abundances of Li, CNO isotopes and $s$-elements are derived.
The derived abundances of $s$-elements nicely agree with theoretical s-process nucleosynthesis 
predictions during the AGB phase. However, the figures obtained for Li and the $^{12}$C/$^{13}$C 
ratios might imply the existence of a non-standard mixing process during the AGB phase operating 
preferentially in low mass stars. The intrinsic or extrinsic nature of C(N) stars is also
discussed. 
\end{abstract}

\section{Introduction}
It is commonly accepted that the spectral sequence along the asymptotic giant
branch (AGB) phase (M$\rightarrow$MS$\rightarrow$S$\rightarrow$SC$\rightarrow$C) is,
in fact, a chemical sequence in which the carbon content in the stellar envelope
continuously increases due to the operation of the so-called 3$^{th}$ dredge-up (TDU)
mechanism. The TDU transports to the envelope fresh carbon immediately 
after each thermal pulse (TP) of the He-shell. In such a way, an AGB star of M-type with
a ratio in the envelope C/O$\sim 0.5$ becomes a carbon C(N) star when this ratio
exceeds unity, C/O$>1$. Another important consequence of TDU is the enrichment of the
envelope in {\it s}-elements. The necessary neutrons for the
$s$-process are released by two reactions: $^{13}$C$(\alpha,n)^{16}$O, which provides the bulk 
of the neutron flux at low neutron densities ($N_n\la 10^7$ cm$^{-3}$), and $^{22}$Ne$(\alpha,n)^{25}$Mg, 
which is activated at temperatures $T\ga 3.0\times  10^8$ K, providing a high peak
neutron density ($N_n\sim 10^{10}$ cm$^{-3}$) and is responsible of the production of $s$-nuclei 
controlled by reaction branchings (see e.g. Busso, Gallino, \& Wasserburg 1999). The abundances of 
$s$-nuclei are known to increase along the above mentioned spectral sequence, as the star gradually ascends 
the AGB. Evidence of this was provided during the last few decades by several studies on AGB stars of
different spectral types (see Busso et al. 2001 and references therein). 

The importance of the study of AGB stars is obvious: i) they are excellent laboratories to test the theory of stellar evolution and nucleosynthesis
(note, in addition, that they are progenitors of planetary nebulae and white dwarfs) ii) in particular
C(N) stars are the main producers of the heavy s-elemnts (A$\geq 90$) in the Galaxy as well as of a substantial fraction
of $^{12}$C, $^{14}$N, $^7$Li and $^{13}$C. However, an evaluation of such 
chemical yields firstly needs accurate abundance determinations. Second, in order to understand the role 
played by these stars in the
chemical evolution of the Galaxy it is mandatory to know their typical masses and the frequency of binarity
among them. In this work, we will address these questions.

\section{Observations and Analysis}
So far the only available abundances were still those by Utsumi (1985) 
based on photographic plates and/or low resolution spectra. C(N) stars are very difficult to analyse spectroscopically 
because of their very crowded spectra. The spectrum of these stars is dominated by molecular absorptions (CN, C$_2$, CH...), 
which require the knowledge of accurate spectroscopic 
information of these molecules for the chemical analysis. The use of high resolution and high signal to noise ratio 
spectra is also mandatory and
even in that case, only a few spectral {\it windows} are available for abundance analysis: $\lambda 4750-4950$ {\AA},
for $s$-elementsand metals, $\lambda\sim 7800$ {\AA} for the Rb analysis, $\lambda 8000-8050$ {\AA} for the
$^{12}$C/$^{13}$C ratios and several Li I lines. In our analysis of a sample of $\sim40$ galactic C(N), we used 
high resolution ($\lambda/\Delta\lambda=4\times 10^4-2\times 10^5$) spectra obtained with the 4.2 m WHT at  
El Roque de los Muchahos and with the 2.2 m telescope at Calar Alto. We use the spectrum synthesis technique to derive
all the abundance figures. The determination of the stellar parameters was made as in Abia et al. (2001) (see this paper
for details).

\section{Results}
\subsection{S-elements}
C(N) stars show overabundances in s-process elements. For the low atomic mass number (Sr,Y,Zr) $s$-elements(ls) we found
a mean enhancement with respect to the metallicity [$<$ls$>$/M]=$+0.70\pm 0.20$, while for the high atomic mass number (Ba,La,Nd,Ce,Sm)
$s$-elements(hs) the mean overabundance is [$<$hs$>$/M]=$+0.52\pm 0.29$. These enhancements are significantly lower than
the previous estimations by Utsumi (1985)\footnote{For an explanation of the differences with respect to Utsumi's data see Abia et al.
(2001).} and set the s-element overabundances in C(N) stars at the same level or slightly higher than those found in S stars.
From the theoretical point of view, this result can be easily understood considering that the overwhelming majority of the C(N) 
stars analysed here have a C/O ratio very close to the unity ($<$C/O$>=1.05\pm 0.02$): in a single episode of TP and TDU, a S-star 
(C/O$\sim 0.9$) can become a C(N) star without increasing significantly the s-element enhancement in the envelope. C(N) stars with
larger C/O ratios and s-element overabundances must lokely exist, however they escape optical detection because of the formation
of a thick circumstellar envelope due to high mass-loss rates. Our C(N) star sample is probably observationally biased in this
sense. In Figure 1 we compare the [hs/ls] vs. [M/H] ratios obtained with theoretical s-process nucleosynthesis models in a
representative case of a 1.5 M$_\odot$ AGB star (see Gallino et al. 1988; Busso et al. 2001 for details). Solid line shows the 
prediction corresponding to a standard (ST) case which assumes the radiative burning of $4\times 10^{-6}$ M$_\odot$ of $^{13}$C in 
a tiny {\it pocket} in the intershell region. The upper and lower curves (dashed lines) limit the region allowed by the models 
according to the different $^{13}$C-pocket choices scaled to the ST case. Model predictions and observations are in nice agreement. Note that the Utsumi's data are marginally in agreement with theoretical models.

\begin{figure}
\plotone{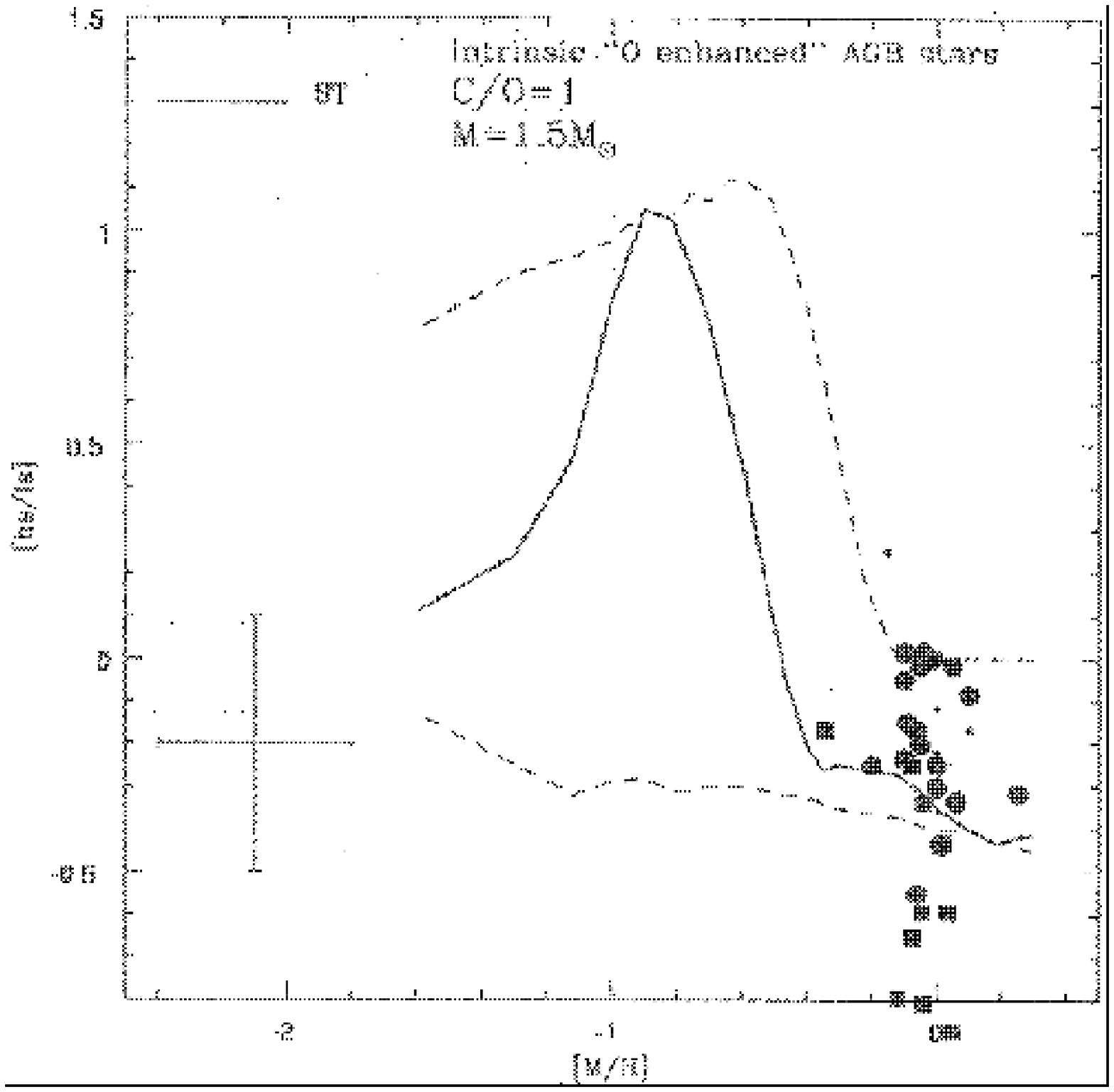}
\caption{The observed (solid circles) mean [hs/ls]  ratio (signature  of the  neutron exposure) as compared with theoretical predictions for a 1.5 M$_\odot$ TP-AGB star. Solid squares: Utsumi (1985). Data points with smaller symbols 
refer to {\it difficult stars}. An initial oxygen enhancement given by 
[O/Fe]$=-0.4$[Fe/H] was assumed in the theoretical models (see Abia et al. 2002 for details).}
\end{figure}

\subsection{The mass}  
Theoretical predictions for s-process as those showed in Figure 1 do not significantly differ for a 3 $M_\odot$. Remarkable
differences exist for more massive stars, but they are not sufficient to allow a clear discrimination of the stellar mass. To 
infer the mass of C(N) stars we can make use of a neutron density sensitive element: Rb.
Depending of the neutron density in the s-process (i.e. depending on whether the $^{13}$C (M$\leq 3$ $M_\odot$) or $^{22}$Ne 
(M$\geq 4$ $M_\odot$) neutron sources provides the bulk of neutrons), a very different abundance pattern between Rb and its 
neighbours (Sr,Y and Zr) is expected. Figure 2
shows the result of such a comparison. It is evident that the predictions for a 1.5 $M_\odot$ model fit much better
the observed [Rb/X] ratios and we can conclude that the majority of C(N) stars studied here are low-mass stars (M$\leq 3$ $M_\odot$).
This figure concerning the most probable mass for C(N) stars is reinforced from the $^{12}$C/$^{13}$C ratios derived. A large
number of the carbon stars studied here have $^{12}$C/$^{13}$C$<40$. These ratios cannot be explained by the standard evolutionary
models on the AGB. Such low ratios might be explained if an extra-mixing process is operating 
during the AGB phase preferably in low mass stars. This non-standard mixing process, perhaps induced by rotation, has been 
named {\it cool bottom processing} (Wasserburg, Boothroyd \& Sackmann 1995; Nollet, Busso \& Wasserburg 2003) and might explain the 
low $^{12}$C/$^{13}$C ratios found in many low mass RGB stars. 

\begin{figure}
\plottwo{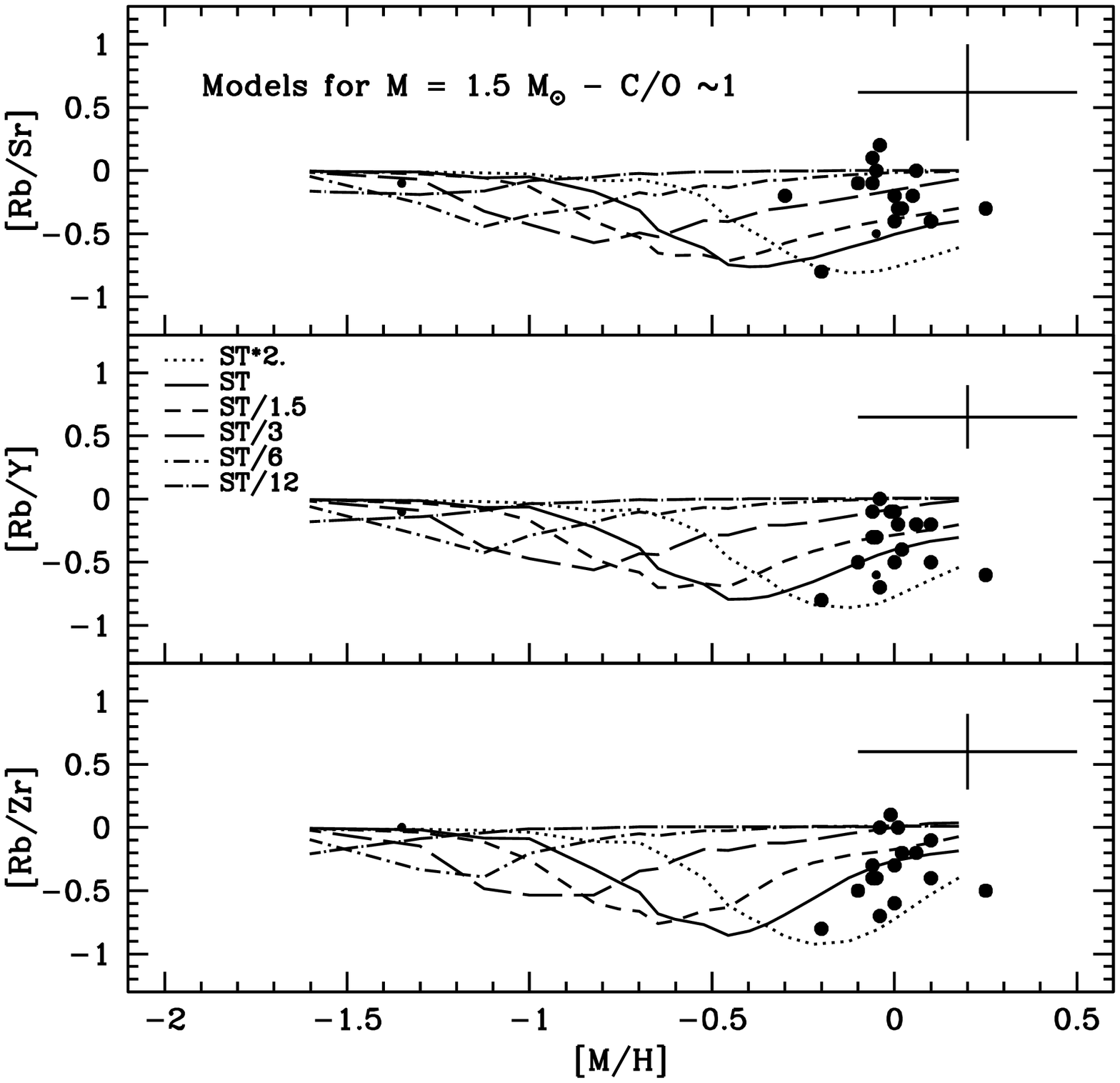}{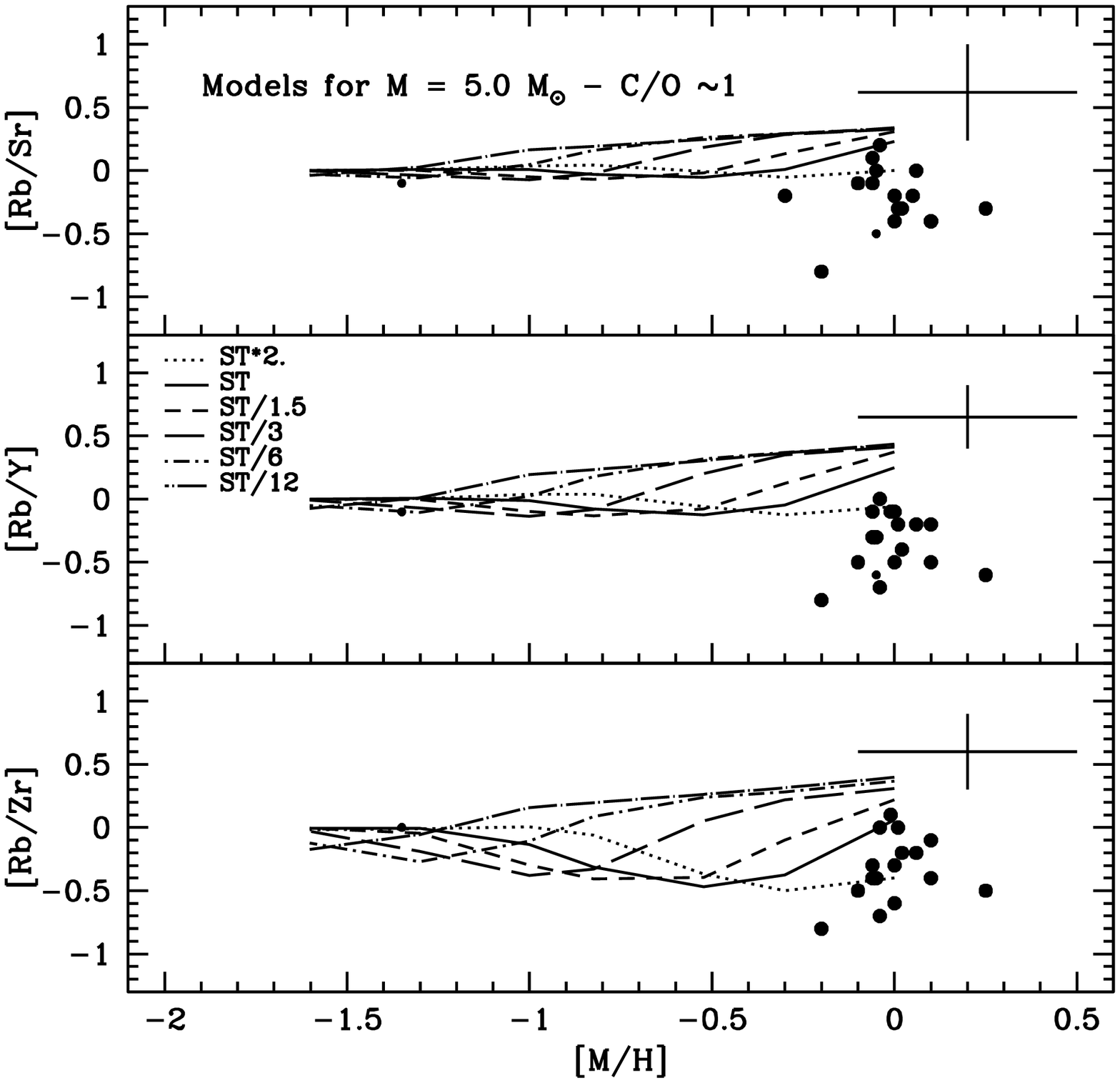}
\caption{Comparison of the observed [Rb/Sr,Y,Zr] ratios vs. [M/H] with different theoretical predictions for a 1.5 and 5 $M_\odot$ 
TP-AGB star when the envelope reaches C/O $\geq$ 1.}
\end{figure}

\subsection{Lithium}
Abia et al. (1993) derived Li abundances in a sample of 230 galactic carbon stars. The most important result of this
work was the confirmation that a small fraction (2-3$\%$) are super Li-rich carbon stars, showing Li abundances
as high as log $\epsilon$(Li)$\sim 5$. In addition, a significant number ($\sim 10\%$) can be considered as Li-rich C-stars, 
with Li abundances in the range 1-2. When interpreting these observations on the basis of theoretical models some problems appear. 
First, how reliable are the Li abundances derived? Yakovina, Pavlenko \& Abia (2002) have recently derived Li abundances in 
super Li-rich C-stars 
using four different Li lines. They show that even considering corrections by N-LTE effects the Li abundances derived from different 
lines show a high dispersion. They  interpreted this as an evidence of the difficulty and uncertainties still present in 
the modelling 
of the atmospheres of AGB stars and put a note of caution when estimating the Li yield from AGB stars. Other figures are also
difficult to understand: 
i) At solar metallicity, the super Li-rich phenomenon is expected for O-rich AGB stars, not for C-rich objects. 
ii) Some super Li-rich stars are fainter (M$_{bol}>-6$) than theoretically predicted. iii) Theoretical models predict maximum 
Li abundances $\sim 4$, however a few stars show peak Li abundances close to 5 (even considering the large error bars). iv) 
How can we explain the large fraction of Li-rich C-stars? Some of these stars are classified as J-type carbon stars which might 
not be AGB stars, but a significant number of them are indeed normal (N) carbon stars. 
v) Why many galactic O-rich and Li-rich AGB stars do not show s-process element enhancements 
(Garc\'\i a-Lario et al. 1999)? Considering that most of Galactic C-stars are of low mass, for which no Li production is
predicted (see Sackmann \& Boothroyd 1992), it seems evident that also an additional non-standard mixing process affecting Li 
might play a role in low mass AGB stars. Theoretical modelling is therefore required, but models like those
discussed by Nollet, Busso \& Wasserburg (2003) can explain in a unique scenario high Li abundances and low $^{12}$C/$^{13}$C
ratios, also predicting anomalies in the oxygen isotopic ratios and consistent production of $^{26}$Al.

\subsection{Intrinsic or Extrinsic C-stars}
Let us finally address the question whether the stars studied here are intrinsic or extrinsic AGB stars, ie: their abundances are 
locally produced by the occurrence of TP during the AGB phase, or generated by mass transfer in binary systems.
As far as we know no significant radial velocity variations have been detected in any of them. Furthermore, $\sim 60\%$ of the sample 
stars show $^{99}$Tc, an incontestable signature of the TP and TDU operation, thus, of their intrinsic nature. A further test can be 
done looking for infrared excess as has been done by Jorissen et al. (1993) in S-stars. If the stars are TP-AGB stars 
they should be high mass-losing stars; the formation of dust and, in consequence, infrared excess are expected. This excess can be measured with the 
flux ratio R$=$F(12 $\mu$m)/F(2.2 $\mu$m). High mass-losing stars (ie. TP-AGB-stars) typically show R$>0.1$. The large majority of 
our sample stars have R$>0.1$, Tc-yes and Tc-no stars being indistinguishable in the R parameter. We might conclude 
that probably all carbon stars in our sample are in fact of intrinsic nature and when they are classified as no-Tc stars, this
is just a consequence of the difficult and uncertain analysis of the Tc I $5924$ {\AA} blend on which this study is based.


\begin{references}
\reference Abia C., Boffin, H.M.J., Rebolo, R., \& Isern, J. 1993, \aap, 272, 455
\reference Abia, C., Busso, M., Gallino, R., Dom\'\i nguez, I.,  Straniero, O., 
\& Isern, J.  2001, \apj, 559, 1117
\reference Abia, C., Dom\'\i nguez, I., Gallino, R., Busso, M., Masera, S., Straniero, O., de Laverny, P., Plez, B., \&
Isern, J. 2002, \apj, (in press)
\reference Busso, M., Gallino, R., \& Wasserburg, G.J. 1999, \araa, 37, 239
\reference Busso,  M.,  Lambert,  D.L., Gallino, R., Travaglio, C., \& Smith, V.V. 2001, \apj, 557, 802
\reference Gallino,  R.,  Arlandini, C., Busso, M., Lugaro,  M., Travaglio, C., Straniero, O.,  Chieffi,
A., \& Limongi, M. 1998, \apj, 497, 388
\reference Garc\'\i a-Lario, P., D'Antona, F., Lub, J., Plez, B., \& Habing, H.J. 1999, in AGB stars, IAU Symposium
N. 191, T. Le Bertre, A. Lebre, C. Waelkens eds., p. 91
\reference Jorissen, A.,  Frayer, D.T., Johnson, H.R., Mayor, M., \& Smith, V.V. 1993, \aap, 271, 463
\reference Nollett, K. M., Busso, M., \& Wasserburg, G. 2003, \apj, (in press)
\reference Sackmann, I.J., \& Bootroyd, A. 1992, \apj, 392, L71
\reference Utsumi,  K. 1985,  in Cool  Stars with Excesses of Heavy Elements, eds. M. Jascheck \& P.C.Keenan
(Dordrecht: Reidel), 243 
\reference Yakovina, L.A., Pavlenko, Y.V., \& Abia, C. 2002, \apss, (in press)
\reference Wasserburg, G.J., Boothroyd, A.I., \& Sackmann, I.J. 1995, \apj, 440, L101
\end{references}
\end{document}